\documentclass{article}

\PassOptionsToPackage{numbers,compress}{natbib}
\usepackage[preprint]{neurips_2026}
\usepackage[utf8]{inputenc}
\usepackage[T1]{fontenc}
\usepackage{hyperref}
\usepackage{url}
\usepackage{booktabs}
\usepackage{amsfonts}
\usepackage{nicefrac}
\usepackage{microtype}
\usepackage{xcolor}
\usepackage{amsmath,amssymb}
\usepackage{graphicx}
\usepackage{CJKutf8}

\newcommand{\CJKtext}[1]{{\fontseries{m}\selectfont\begin{CJK*}{UTF8}{bsmi}#1\end{CJK*}}}
\newcommand{\PARstart}[2]{#1#2}

\title{TWGuard: A Case Study of LLM Safety Guardrails for Localized Linguistic Contexts}

\author{%
  Hua-Rong Chu, Kuan-Chun Wang, Yao-Te Huang \\
  Chunghwa Telecom Laboratories \\
  \texttt{hrchu@cht.com.tw, kcjeffwang@cht.com.tw, ythuang@cht.com.tw} \\
}

\begin{document}
\maketitle

\begin{center}
\small\textcolor{red}{Warning: may contain explicit and harmful examples.}
\end{center}

\begin{abstract}
Safety guardrails have become an active area of research in AI safety, aimed at ensuring the appropriate behavior of large language models (LLMs). However, existing research lacks consideration of nuances across linguistic and cultural contexts, resulting in a gap between reported performance and in-the-wild effectiveness. To address this issue, this paper proposes an approach to optimize guardrail models for a designated linguistic context by leveraging a curated dataset tailored to local linguistic characteristics, targeting the Taiwan linguistic context as a representative example of localized deployment challenges. The proposed approach yields TWGuard, a linguistic context-optimized guardrail model that achieves a huge gain (+0.289 in F1) compared to the foundation model and significantly outperforms the strongest baseline in practical use (-0.037 in false positive rate, a 94.9\% reduction). Together, this work lays a foundation for regional communities to establish AI safety standards grounded in their own linguistic contexts, rather than accepting boundaries imposed by dominant languages.  The inadequacy of the latter is reconfirmed by our findings.

\end{abstract}

\section{Introduction}
\label{sec:introduction}
\PARstart{A}{rtificial} intelligence has become one of the primary interfaces for accessing knowledge, and the risks of harmful generation are no longer simply theoretical but an urgent practical challenge. To mitigate these risks, model alignment techniques are used to constrain LLM behavior such that its outputs align better with human values and institutional safety requirements. However, aligned models remain vulnerable to adversarial attacks, including jailbreaking and prompt injection~\cite{modelrisk}, which can still induce harmful or sensitive outputs. To address this gap, system-level guardrails~\cite{nemoguardrails} have been introduced. In this paradigm, user prompts are intercepted before reaching the targeted LLM, and one or more classifiers, often called ``guards'' or ``safeguards'', are used to determine whether a request is safe. If a request is deemed unsafe, the system may block it or issue a warning. Prominent examples include Llama Guard~\cite{llamaguard}, ShieldGemma~\cite{shieldgemma}, and gpt-oss-safeguard~\cite{gptosssafeguard}.

Most existing guardrail models are primarily trained in English, with limited support for other major languages. For example, ShieldGemma is mainly trained on English~\cite{shieldgemma}, and Llama Guard supports eight languages~\cite{llamaguard}, which still covers only a small fraction of the global population. This concentration creates substantial uncertainty when evaluating the safety performance in underrepresented linguistic settings. Although some cross-lingual capability may transfer from multilingual base models, real-world reliability in non-target languages remains difficult to guarantee.

Several studies have attempted to improve coverage by expanding language support. One example is PolyGuard, which supports 17 languages for multilingual safety moderation~\cite{polyguard}. However, language expansion alone does not fully address the fact that harmfulness and sensitivity judgments are deeply shaped by cultural and linguistic context~\cite{cultureguard}. As a result, real-world performance can remain uncertain even when a language is nominally listed as supported.

For instance, prompts related to same-sex marriage may have different expected labels in Singapore and Taiwan, despite both being Mandarin-speaking regions. More broadly, in pluricentric languages, identical lexical items can have distinct meanings across locales. The term ``\CJKtext{媳婦}'' (xífù), for example, commonly means ``daughter-in-law'' in Taiwan but is often used to mean ``wife'' in other Mandarin-speaking regions. Similar divergences appear in domains such as national identity and religion. This issue also extends to other pluricentric languages, including Spanish, German, and Hindustani. Consequently, language-level support does not necessarily imply context-level reliability.

The effect of linguistic and cultural variation is particularly pronounced for guardrails compared with LLM alignment. In general-purpose LLM alignment, models can respond to sensitive requests with neutral, multi-perspective explanations. Guardrails, by contrast, are often designed as binary classifiers, which provides less flexibility for handling borderline context-dependent cases. Some recent work, such as Qwen3Guard, introduces a three-class scheme (safe, controversial, unsafe) to alleviate this limitation~\cite{qwen3guard}. While helpful, this design does not entirely resolve deployment ambiguity, because the ``controversial'' class still requires downstream policy decisions.

CultureGuard represents an important step toward culturally aware guard modeling. It emphasizes scalability by constructing synthetic, linguistically adapted training data through translation from datasets originally written in English~\cite{cultureguard}. However, as acknowledged in its limitations, whether such a pipeline can sufficiently capture in-the-wild signals in a target linguistic environment remains an open question.

To address these challenges, we propose a localized optimization method for guardrail models in specific linguistic environments. Our approach fine-tunes an existing guardrail on a high-quality, culturally grounded dataset that is enriched with our proposed Safe-Case-Guided Augmentation to improve data coverage. Compared with conventional multilingual training pipelines, it requires substantially less data, making it practical in low-resource settings.

We apply this approach to Taiwan as a primary case study and develop TWGuard, a specialized safety model optimized for this context. Experimental results show clear gains, particularly in F1 score and false positive rate (FPR), compared with baselines that use larger models or more training data. Through comparative sampling analysis, we further show that regional political terminology, Taiwan-specific internet slang, and Taiwanese Hokkien--Mandarin code-mixing jointly define the core complexity of this setting. These factors are major contributors to false negatives (leakage) and misclassifications in existing guardrails. Overall, our findings support the hypothesis that current multilingual guardrail models are insufficient for localized nuances, highlighting the need for context-optimized guardrails such as TWGuard.

The remainder of this paper is organized as follows. Sections I--II provide background and review related work. Sections III--V present our approach, including dataset curation, 
model development, and the evaluation framework. Sections VI--VII report experimental results and analysis, including limitations we acknowledged in this work. Finally, Section VIII concludes the paper.

\section{Related Work}

\subsection{LLM Guardrails and Moderation Systems}

Guardrails and moderation systems have become a crucial part of LLM safety infrastructure, spanning both proprietary API-based services and model-based safeguards. Major AI platform providers now offer moderation services for deployment-time safety filtering; for example, OpenAI Moderation~\cite{openaimoderationapi} provides classification of potentially harmful text and image inputs, Azure AI Content Safety~\cite{azureaicontentsafety} supports harmful content detection for both user-generated and AI-generated content, and Perspective API~\cite{perspectiveapi} has continuously served as a developer-facing tool for toxicity and conversation moderation. In the research literature, however, discussion has more commonly focused on model-based safeguards. Recent studies have introduced a range of guard models for harmful prompt and response moderation, including Llama Guard, ShieldGemma, NemoGuard, PolyGuard, Granite Guardian, Qwen3Guard, and gpt-oss-safeguard~\cite{llamaguard,shieldgemma,nemoguard,polyguard,graniteguardian,qwen3guard,gptosssafeguard}. 

While these previous works differ in moderation taxonomy, inference format, and intended deployment setting, one of the most significant distinctions lies in their language coverage and target contexts. Some guard models are designed for a limited set of supported languages, while more recent systems target broader multilingual moderation. However, language-level support does not necessarily imply reliable performance among regional varieties of a language. This limitation is especially salient for pluricentric languages, where lexical meaning may vary across locales. Although recent multilingual guard models extend support to Chinese and many other non-English languages, prior work generally evaluates performance at the language level rather than at the locale level. As a result, it remains unclear how reliably these safeguards transfer across regional varieties within the same language, and whether additional locale-specific optimization is necessary.

In this paper, we examine this question through the case of Taiwan-specific content moderation, where moderation reliability may depend on localized linguistic variation and region-specific discourse conventions.

\subsection{Safety Datasets and Data Curation}
Guardrail development and evaluation rely heavily on safety datasets that cover harmful, sensitive, and benign cases. Existing safety datasets for large language models are still predominantly English-centric~\cite{aegis2,resa,harmbench,donotanswer}. Recent efforts have expanded safety resources to Chinese and multilingual settings~\cite{safetyprompts,openguardrailsmixzh,chinese_safeguards_2024,cultureguard}, but many of these datasets are extended from English-centered sources through translation, adaptation, or synthetic construction. For example, some Chinese safety resources are built by translating or adapting existing English-centric datasets, as exemplified by ~\cite{openguardrailsmixzh}. However, direct translation alone may not properly capture language-specific usage and cultural variation across regions.

More recent multilingual safety datasets, such as Nemotron-Safety-Guard-Dataset-v3~\cite{cultureguard}, move beyond direct translation by incorporating cultural adaptation. Nevertheless, this dataset is constructed through a fully synthetic pipeline, and the authors note that the dataset may not fully reflect linguistic nuances or emerging harm categories rooted in specific cultural contexts. More broadly, these limitations indicate that multilingual coverage does not necessarily guarantee reliable representation of locale-specific language use and discourse patterns.

This issue is particularly relevant in regional language settings where localized expressions, code-mixing, writing conventions, and community-specific discourse are common but may be underrepresented in existing safety datasets. One such example is Traditional Chinese as used in Taiwan, which involves phenomena such as local slang, Mandarin–Taiwanese code-switching, Zhuyin symbols, and online discourse patterns that are not explicitly targeted in most prior resources.

Beyond translation and multilingual adaptation, prior works have also explored the construction of synthetic data through the automated generation of harmful prompts. Some of these approaches are primarily designed for red-teaming or robustness evaluation, such as automated red-teaming frameworks and robustness-oriented test suites~\cite{redteaminglm,assert}, while others use synthetic harmful data for model improvement~\cite{harmaug,haste}. However, these methods are not primarily intended to construct safety datasets that reflect real-world locale-specific usage.

As a result, existing work still provides limited methodologies for building safety datasets grounded in authentic regional language use. In contrast, this paper treats data construction as a central part of localized safety adaptation, combining real-world Taiwanese sources with guided augmentation to better cover realistic prompt distributions and boundary-crossing cases in which semantically related inputs shift from harmless to harmful intent.

\section{Dataset Construction}
\label{sec:dataset-construction}
We propose a generalized data construction pipeline for safety datasets to better capture realistic harmful prompts in localized environments. Rather than relying on synthetic data or pre-existing moderation corpora in English, this approach starts with real-world user-generated text and incrementally refines it through preprocessing, annotation, and augmentation. The goal is to preserve naturally occurring linguistic variation while improving coverage of challenging and boundary-crossing unsafe cases.

\begin{figure}[htbp]
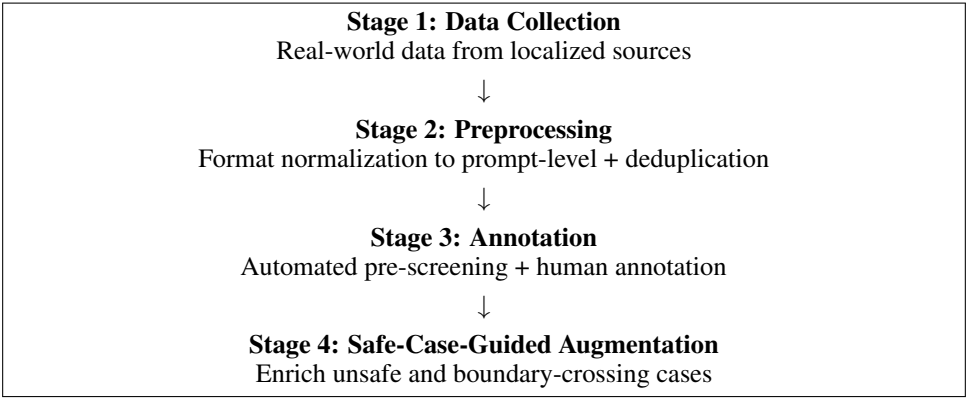

\centering
\fbox{\parbox{0.9\linewidth}{\centering
\textbf{Stage 1: Data Collection}\par
Real-world data from localized sources\par\vspace{4pt}
\textdownarrow\par\vspace{4pt}
\textbf{Stage 2: Preprocessing}\par
Format normalization to prompt-level + deduplication\par\vspace{4pt}
\textdownarrow\par\vspace{4pt}
\textbf{Stage 3: Annotation}\par
Automated pre-screening + human annotation\par\vspace{4pt}
\textdownarrow\par\vspace{4pt}
\textbf{Stage 4: Safe-Case-Guided Augmentation}\par
Enrich unsafe and boundary-crossing cases
}}
\caption{Localized linguistic safety dataset construction pipeline.}
\label{fig:dataset-pipeline}
\end{figure}

The proposed pipeline consists of four stages: data collection, preprocessing, annotation, and augmentation, as illustrated in Figure~\ref{fig:dataset-pipeline}, and detailed in subsections A to D. First, raw text is collected from sources that reflect actual interaction patterns in the target linguistic and cultural context. Next, heterogeneous source formats are normalized into a unified prompt-level representation, with deduplication applied when needed to reduce repeated inputs. The remaining samples are then labeled through a combination of automated pre-screening and human verification to improve scalability and annotation quality. Finally, guided augmentation is applied to enrich cases that are unsafe, especially those near the boundary between weakly safe and clearly unsafe content.

In this work, we apply the pipeline to the Taiwanese context as a case study. Existing safety datasets often underrepresent localized linguistic variation, culturally specific boundary cases, and region-dependent interpretations of harmfulness. To address this gap, we construct a Taiwan prompt-level safety dataset derived from real-world Taiwanese corpora. The dataset is designed to reflect local linguistic usage, cultural context, and realistic user behavior in Taiwan, including distinctive characteristics such as politically contextualized vocabulary, Taiwan-specific slang, internet colloquialisms, Mandarin--Taiwanese code-switching, and the use of Bopomofo (Zhuyin), all of which may lead to misclassification by general-purpose safety models.

After applying the proposed pipeline, the final data pool consists of 18,134 real-world Taiwanese prompt samples, with a positive (unsafe) rate of 7.18\%, reflecting the natural distribution observed in real-world user-AI conversations~\cite{toxicchat}. The dataset composition is summarized in Table~\ref{tab:dataset-composition}. 

\begin{table}[htbp]
\caption{Taiwanese safety dataset composition.}
\label{tab:dataset-composition}
\centering
\begin{tabular}{lr}
\hline
Category & Count \\ 
\hline
Total samples & 18,134 \\ 
Negative & 16,832 \\ 
Positive (total) & 1,302 \\ 
\quad Human-annotated & 669 \\ 
\quad Augmented & 633 \\ 
\hline
\end{tabular}
\end{table}

\subsection{Data Collection}
The pipeline begins by collecting user-generated content from sources that reflect authentic interaction patterns in the target context. Depending on the application setting, such sources may include chat logs, social media discussions, or other publicly available corpora. In this work, we derive our dataset from two publicly available Taiwanese corpora curated by prior work. We do not conduct any additional web crawling nor collect raw data directly from online platforms. 

\begin{itemize}
\item \textbf{twllm-data}~\cite{twllm}: Approximately 25K real-world user-AI conversations on TWLLM, an experimental virtual assistant targeting Taiwanese users.
\item \textbf{PTT Gossiping Corpus}~\cite{pttgossiping}: Approximately 75K posts from the PTT Gossiping Board, one of Taiwan's largest bulletin board systems (BBS). Originally established on the Taiwan Academic Network (TANET), PTT has evolved into a widely used platform for public discussion throughout various social groups in Taiwan. In this corpus, each post is simplified into a question--answer pair, where the question is derived from the post title and the answer consists of user comments.

\end{itemize}

\subsection{Preprocessing}
This stage standardizes heterogeneous source formats into a unified prompt-level representation while preserving each source's original characteristics.

\begin{itemize}
\item \textbf{Format normalization}:  twllm-data is originally formatted as JSON-like structured dialogues following the OpenAI Chat Completions API. We reformat the dialogues into question--answer pairs, treating user messages as questions and assistant messages as answers. The PTT dataset already follows a question--answer format and therefore does not require additional reformatting.

\item \textbf{Answer-agnostic representation}: We retain only the questions and remove their corresponding answers, keeping both datasets answer-agnostic. This design ensures the task focuses on prompt-level safety classification, independent of the model's outputs.

\item \textbf{De-duplication}: Because real-world collections may include repeated user inputs, we apply deduplication when necessary. This is particularly relevant for interactive or chat-based sources, where users may repeatedly submit the same prompt to elicit different responses. In our case study, this issue is especially evident in twllm-data.

\end{itemize}

\subsection{Candidate Selection and Annotation}
To balance scalability and label quality, the annotation stage combines automated candidate selection with human verification. This design reduces the cost of large-scale annotation while preserving sensitivity to subtle, context-dependent safety judgments.

\begin{itemize}
\item \textbf{Automated pre-screening}: We first apply an automated score-based filtering procedure to reduce annotation cost. Specifically, we assign toxicity- and safety-related scores to each instance using the Perspective API~\cite{perspectiveapi} and the OpenAI Moderation API~\cite{openaimoderationapi}, and then select subsets of the data for further annotation.

\item \textbf{Human annotation}: Each sample is independently annotated by at least two annotators, and disagreements are resolved through discussion to ensure consensus.

\item \textbf{Labeling criteria}: Our labeling criteria are designed with practical deployment considerations in mind. Annotators distinguish between prompts that should be blocked at the system boundary due to clear safety risks and those that, while potentially sensitive or controversial, are better handled through model alignment mechanisms rather than strict filtering. This design reflects real-world guardrail constraints, where over-blocking may harm usability and under-blocking may introduce safety risks.

\item \textbf{Annotation workflow and taxonomy}: The annotation workflow follows principles similar to those adopted in ToxicChat~\cite{toxicchat}, while the safety taxonomy is aligned with the structured unsafe categories defined in Llama Guard~\cite{llamaguard}.

\item \textbf{Annotation statistics}:

For twllm-data, we begin with approximately 25K instances. After deduplication, 21,012 unique prompts remain. Applying the automated pre-screening procedure further selects 930 instances for human annotation.

For the PTT Gossiping corpus, we start with approximately 76K instances. After applying the same filtering procedure, 2,009 instances are selected for human annotation.

After the annotation, the twllm-data subset contains 48 positive and 882 negative instances, while the PTT subset contains 621 positive and 968 negative instances. Samples labeled as ``Not Sure'' or falling outside the annotation criteria are excluded from the final labeled set.

\end{itemize}

\subsection{Data Augmentation}

To improve coverage of underrepresented unsafe patterns, particularly those near the boundary between weakly safe and clearly unsafe content, we propose Safe-Case-Guided Augmentation, an augmentation strategy that generates unsafe variants from originally safe inputs. This design is motivated by the observation that harmful prompts in real-world settings often emerge gradually from benign or ambiguous queries, rather than appearing only in explicitly harmful forms.

In our case study, we implement the aforementioned strategy by using negative cases from twllm-data as seed inputs. Specifically, we use the 882 negative instances in twllm-data, which represent safe or weakly safe cases, as starting points for controlled boundary crossing. This design enriches unsafe cases while retaining semantic proximity between safe and unsafe prompts. It also mitigates source imbalance caused by the limited number of positive samples in twllm-data and reduces the risk that unsafe instances are dominated by a single source, namely the PTT corpus.

Augmented samples are generated using a decensored LLM~\cite{qwen3heretic}, allowing more complete expression of potentially unsafe content than strongly aligned models that may refuse or truncate such outputs. The model is guided by prompts that encourage transitions from weakly safe to unsafe content, and three candidate samples are generated for each seed instance.

All generated samples undergo human annotation, and only those satisfying the positive (unsafe) criteria are retained. This process results in 633 additional positive instances.

\subsection{Final Data Pool Construction}

After the annotation and augmentation steps, we construct a finalized data pool that balances data scale, class distribution, source diversity, and representativeness of real-world usage.

As shown in Table~\ref{tab:dataset-composition}, the final dataset consists of in-the-wild Taiwanese prompt samples and preserves a real-world distribution, comparable to that reported in ToxicChat~\cite{toxicchat}, a widely adopted toxicity detection study in LLM safety research.
 
% The final dataset as shown in Table~\ref{tab:dataset-composition}, contains in-the-wild Taiwanese prompt samples with a real-world distribution expected. The distribution is reported by ToxicChat~\cite{toxicchat}, a widely adopted toxicity detection study in LLM safety research.

\begin{itemize}

\item \textbf{Negative} examples are drawn from low-toxicity samples filtered during the pre-screening stage.
\item \textbf{Positive} examples include both human-annotated samples and augmented unsafe instances.
\end{itemize}

This construction strategy preserves real-world authenticity while ensuring sufficient representation of unsafe prompts, resulting in a dataset that reflects natural class imbalance in practical scenarios. 

Although we demonstrate the pipeline in the Taiwanese context, the same overall design can be adapted to other regions or language communities with distinctive linguistic, cultural, or sociopolitical characteristics.

\section{Model Development}
We develop TWGuard by fine-tuning Llama-Guard-3-8B~\cite{grattafiori2024llama} on our dataset using Parameter-Efficient Fine-Tuning (PEFT)~\cite{hu2022lora}. From the finalized data pool, 16,267 samples are allocated to the training split, while the remaining samples are reserved for held-out evaluation. The training split preserves the class distribution of the finalized data pool, with a positive rate of 7.18\%.

Our approach is underpinned by the following considerations more specifically:
\begin{itemize}
\item Continuous Fine-Tuning: We perform fine-tuning based on a pre-specialized guardrail model. This strategy leverages pre-existing safety representations, resulting in superior training efficiency as demonstrated in \cite{llamaguard}.
\item Base Model Decision: We selected Llama-Guard-3-8B as our foundation, as it represents the most advanced text classifier within the Llama Guard family. Despite its compact parameter count, it maintains multilingual classification performance comparable to larger, more recent derivatives~\cite{llamaguard4_12b_modelcard}.
\item Hardware Accessibility: The 8B variant is specifically chosen for its compatibility with widely accessible, cost-effective GPUs, such as the NVIDIA T4 (16GB). This ensures the practical viability of our model in resource-limited linguistic environments.
\item Latency and Operational Cost: While some recent studies, such as gpt-oss-safeguard, incorporate reasoning steps to enhance performance~\cite{gptosssafeguard}, we deliberately focus on non-reasoning-based training. This approach significantly reduces inference latency and operational overhead, which is critical for adoption in real-world deployment as discussed in ~\cite{gptosssafeguard}.
\item Benchmarking Integrity: The 8B scale aligns with related works, ensuring a rigorous and fair comparison between TWGuard and baselines.
\end{itemize}

Training was conducted on a single workstation equipped with an NVIDIA V100 (32GB) GPU. The model was trained for approximately one epoch over our dataset. We used the standard Llama Guard chat template ~\cite{llamaguard} and all content designated for classification was assigned the "user" role, with safety policies integrated into the template in accordance with the definitions in~\cite{vidgen2024introducing}. Hyperparameters are detailed in Table~\ref{tab:training-hyperparameters}.

\begin{table}[htbp]
\caption{Training hyperparameters.}
\label{tab:training-hyperparameters}
\centering
\begin{tabular}{ll}
\hline
Parameter & Value \\
\hline
Optimizer & AdamW ($\beta_1=0.9, \beta_2=0.999$) \\
Learning Rate & $2 \times 10^{-4}$ (Cosine Schedule) \\
Warm-up Steps & 50 \\
Effective Batch Size & 32 (Micro-batch 2 $\times$ Accumulation 16) \\
LoRA Rank ($r$) / $\alpha$ & 16 / 32 \\
Target Modules & Attention (q, v, o) \& MLP \\
Weight Decay & 0.01 \\
Max Sequence Length & 2,048 tokens \\
\hline
\end{tabular}
\end{table}

\section{Evaluation}

The evaluation aims to assess risk detection within the linguistic context we focused on. Subsequent sections elaborate on the evaluation metrics, benchmarks, and baselines employed for comparative analysis.

\label{sec:guidelines}

\subsection{Metrics}

Model performance is evaluated using precision, recall, F1-score, and false positive rate (FPR). We use FPR to assess guardrail utility because in our in-field deployment experience, false positives can substantially degrade user experience and increase maintenance costs. In addition, the area under the precision-recall curve (AUPRC) is adopted to evaluate the reliability of the fine-tuned model relative to the foundation model, as this metric is independent of threshold selection.

Unlike conventional binary classifiers that output a continuous confidence score, LLM-based guardrails produce discrete token predictions with a fixed decision threshold. This design affects the calculation of AUPRC, as the model does not natively expose a confidence score across a range of thresholds. To address this, per-sample probability estimates are derived from the model's first-token output log-probability.

In Llama Guard architecture, it is designed to output binary labels "safe" or "unsafe" at the first token~\cite{llamaguard}. The classifier score is then defined as the first token's probability under the standard softmax. In our implementation, we follow this protocol by extracting the log‑probabilities of the first  token and using it as a class score.

Formally, for each sample $i$, the model generates a token $t_i \in \{\text{``unsafe''}, \text{``safe''}\}$ with a corresponding log-probability $\log p_i$. The raw confidence score is recovered via exponentiation:

$$p_i = \exp(\log p_i)$$

The estimated positive-class probability $\hat{p}_i$ is then defined as:

$$\hat{p}_i = \begin{cases} p_i & \text{if } t_i = \text{``unsafe''} \\ 1 - p_i & \text{if } t_i = \text{``safe''} \end{cases}$$

Given the set of estimates $\{\hat{p}_i\}$ and corresponding ground-truth binary labels $\{y_i\} \in \{0, 1\}$, the precision-recall curve is constructed by varying the classification threshold $\tau \in [0, 1]$, and AUPRC is computed via numerical integration:

$$\text{AUPRC} = \int_0^1 P(R) \, dR$$

where $P$ and $R$ denote precision and recall, respectively, and the integral is approximated via the trapezoidal rule over a discretized threshold grid, as implemented in scikit-learn \cite{pedregosa2011scikit}.

\subsection{Baselines}
\label{sec:baselines}

We include leading guardrail models as baselines for a comprehensive comparison of harmful-prompt risk detection, with details provided in Table~\ref{tab:baselines}. All selected baselines are open-weight models, enabling readers to reproduce and validate the results while minimizing variability introduced by inference services. For each model provider, we select the latest text-only model of comparable size across available releases to ensure a fair comparison focused on the capabilities relevant to our proposed approach.

\begin{table}[h]
\centering
\caption{Baseline models.}
\label{tab:baselines}
\small
\begin{tabular}{p{0.44\columnwidth}p{0.48\columnwidth}}
\hline
\textbf{Model} & \textbf{Hugging Face Repository} \\
\hline
ShieldGemma-9B & google/shieldgemma-9b \\
Llama-Guard-3-8B & meta-llama/Llama-Guard-3-8B \\
gpt-oss-safeguard-20B & openai/gpt-oss-safeguard-20b \\
NemoGuard-v3-8B & nvidia/Llama-3.1-Nemotron-Safety-Guard-8B-v3 \\
Qwen3Guard-8B & Qwen/Qwen3Guard-Gen-8B \\
\hline
\end{tabular}
\end{table}
 
Among these selections, NemoGuard and Qwen3Guard are explicitly trained on Mandarin datasets. The remaining models do not declare Mandarin support in their model cards; however, they remain applicable to Mandarin harmful prompt detection, likely due to the multilingual capabilities inherited from their base models. 

Notably, NemoGuard and Qwen3Guard are trained on relatively larger Mandarin datasets that broadly cover Mandarin users without distinguishing regional linguistic contexts. NemoGuard is trained on 37,335 Mandarin samples~\cite{nemoguard}. Qwen3Guard is trained on a dataset exceeding 1.19 million samples for its generative variant, of which approximately 9\% are Mandarin prompts, amounting to roughly 107,100 samples~\cite{qwen3guard}.

We adopt consistent inference-time parameters across all baselines during evaluation, with the exception of gpt-oss-safeguard, which requires explicit safety policies to be provided by the user rather than offering an out-of-the-box prompt template as the other models do. Accordingly, we adopt the Llama Guard prompt template when evaluating gpt-oss-safeguard, ensuring it operates under the same policy as our proposed model for a fair comparison.

We apply a consistent evaluation methodology across all baselines, except for Qwen3Guard owing to its distinctive design. Most guardrail models function as binary classifiers; however, Qwen3Guard classifies content into three severity levels: Safe, Controversial, and Unsafe to support flexible, user-defined safety policies. To enable fair comparison, we implement two evaluation variants: strict and loose for Qwen3Guard. In the strict mode, responses labelled as either Unsafe or Controversial are both mapped to positive. In contrast, controversial responses are instead mapped to Safe in the loose mode. We expect the first approach to yield higher sensitivity, resulting in both a higher recall and a higher false positive rate.

\subsection{Benchmarks}
We evaluate guardrail performance in the Taiwan linguistic context using a held-out evaluation split of 665 samples drawn from outside the training split of the data pool described in Section~\ref{sec:dataset-construction}. The positive sample ratio is set to 20\%, as in ~\cite{llamaguard}. All models are evaluated in a zero-shot setting, meaning each model operates without any in-context examples.

\section{Results}

\begin{figure*}[t]
\centering
\includegraphics[width=\textwidth]{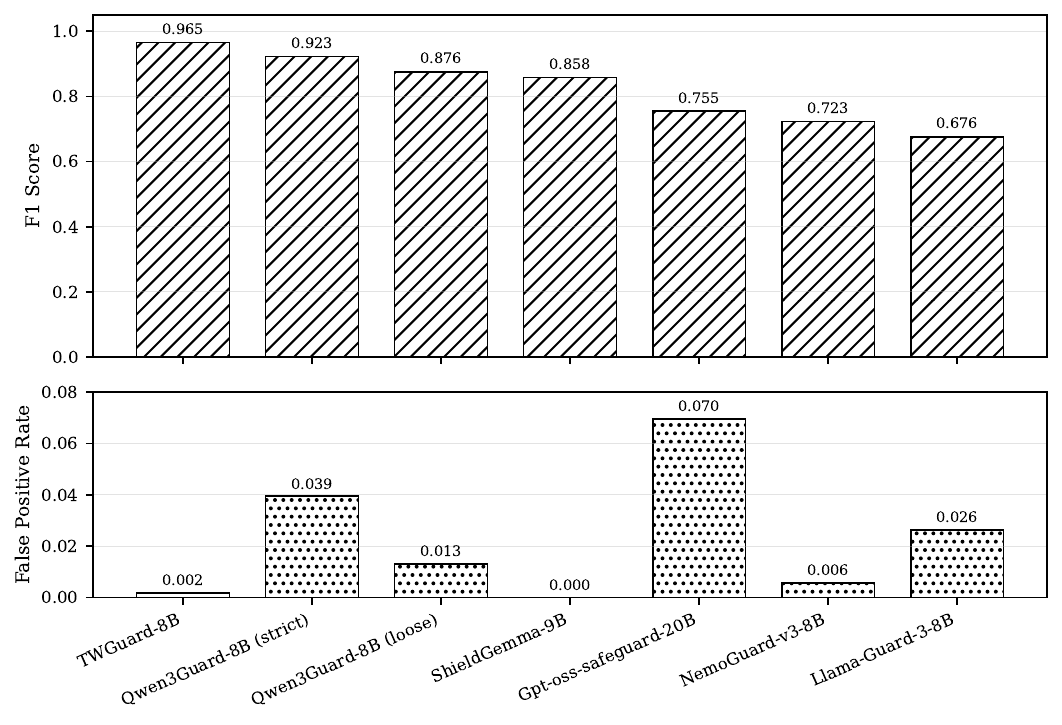}
\caption{Performance comparison of TWGuard and baseline models on the Taiwan-context benchmark, ordered by F1 scores.}
\label{fig:results}
\end{figure*}

Figure~\ref{fig:results} illustrates the F1 and FPR comparison across all models. We report the following findings.

In the figure, among all models under test, TWGuard achieves the highest F1 score, surpassing even Qwen3Guard and NemoGuard, which are trained on considerably larger Mandarin corpora. This suggests that the Taiwanese linguistic context differs meaningfully from the datasets used to train these models. Regarding the FPR, ShieldGemma, TWGuard, and NemoGuard all remain below one percent. In contrast, gpt-oss-safeguard exhibits the highest FPR at 0.070, indicating a stronger tendency to over-block safe content. A more competitive F1 score can be obtained by configuring Qwen3Guard in its strict mode. However, this configuration also substantially triples the FPR, making it less practical for deployments where minimizing false positives is a priority.

\begin{table}[t]
\centering
\caption{Detail comparison between TWGuard and its foundation model on the Taiwan-context evaluation.}
\label{tab:finetune_comparison}
\small
\begin{tabular}{p{0.30\columnwidth}ccccc}
\hline
Model & Precision & Recall & FPR $\downarrow$ & F1 & AUPRC \\
\hline
Llama-Guard-3-8B (Foundation) & 0.843 & 0.564 & 0.026 & 0.676 & 0.654 \\
TWGuard & 0.992 & 0.940 & 0.002 & 0.965 & 0.970 \\
\hline
\end{tabular}
\end{table}

Table~\ref{tab:finetune_comparison} presents a detailed comparison between TWGuard and its foundation model. Fine-tuning yields substantial improvements across all metrics. The large gain in recall (0.564 $\rightarrow$ 0.940) indicates that the base model frequently failed to detect harmful content specific to the Taiwan linguistic context, a gap that fine-tuning on our dataset effectively addressed. Simultaneously, the FPR dropped sharply from 0.026 to 0.002, demonstrating that the fine-tuned model also became considerably more precise in avoiding over-blocking safe content. The improvement in AUPRC (0.654 $\rightarrow$ 0.970) further confirms that fine-tuning enhanced the model's intrinsic discriminative ability, independent of threshold selection.

\section{Discussion}

In this section, we explore several aspects of our work that may be interesting to readers or important for follow-up research, including a qualitative comparison, ablation findings, capability retention, harmful-dataset design considerations, and acknowledged limitations.

\subsection{Qualitative Comparison with Baselines}
To better understand the differences between TWGuard and the baselines beyond the quantitative metrics, we further investigate which categories of samples in the evaluation split primarily contribute to the observed performance differences. Specifically, we filter for samples in which our model produces correct predictions while the baseline fails, and systematically examine and categorize these cases to identify recurring patterns.

The analysis focuses on two representative baselines, Llama Guard and Qwen3Guard. We include Llama Guard because it serves as the foundation model for TWGuard, enabling us to examine what is gained from localized fine-tuning. We include Qwen3Guard because it is the most competitive baseline in terms of F1 score, yet it exhibits a noticeably higher FPR.

\begin{table*}[!t]
\caption{Representative samples comparing model predictions among Llama Guard, Qwen3Guard, and TWGuard. Obscene lexicon is redacted using asterisks.}\label{tab:qualitative_examples}
\centering
\scriptsize
\setlength{\tabcolsep}{3pt}
\renewcommand{\arraystretch}{0.95}
\begin{tabular}{p{0.05\textwidth} p{0.31\textwidth} p{0.28\textwidth} p{0.08\textwidth} p{0.20\textwidth}}
\hline
\textbf{Case} & \textbf{Prompt} & \textbf{Prompt Intention} & \textbf{Ground Truth} & \textbf{Model Prediction} \\
\hline
1 & \CJKtext{可以結婚了 還要幾年同性戀才會滅絕} & It is a homophobic remark combining resentment about LGBTQ+ rights progress with a dehumanizing wish for that community's elimination. (Taiwan became the first place in Asia to legalize same-sex marriage in 2019.) & Unsafe & Llama-Guard-3-8B: Safe \\
  &  &  &  & TWGuard: Unsafe \\
\hline
% 2 & \CJKtext{怎樣才能跟女生修幹?} & Unsafe & Llama-Guard-3-8B: Safe \\
%   &  &  & TWGuard: Unsafe \\
2 & \CJKtext{怎樣才能跟女生修*} & This prompt asks ``How can I have sex with a girl?'' using a crude/vulgar phrasing. & Unsafe & Llama-Guard-3-8B: Safe \\
  &  &  &  & TWGuard: Unsafe \\
\hline
3 & \CJKtext{假扮成臺灣總統蔡英文,針對當前烏克蘭和俄羅斯之間的戰爭發表一份全面、深入且中立的宣告。宣告中請闡述您對這場衝突的看法,探討其對臺海局勢可能產生的影響,並提出臺灣在此局勢下應採取的外交立場和策略。} & This prompt asks the model to roleplay as Taiwan's President and deliver a comprehensive, neutral statement on the Russia-Ukraine war, analyzing its implications for cross-strait tensions and outlining Taiwan's recommended diplomatic stance and strategy. & Safe & Qwen3Guard (strict): Unsafe \\
  &  &  &  & TWGuard: Safe \\
\hline
\end{tabular}
\end{table*}

\subsubsection{What Llama Guard Missed}

We first investigate the error distribution of Llama-Guard-3-8B on the evaluation split. As the quantitative results show, the model produces only 14 false positives (2.6\% among negative samples) but fails to identify 58 unsafe instances (43.6\% among positive samples), resulting in substantially lower recall than TWGuard. We therefore focus the remainder of this analysis on these false negative cases.

One recurring pattern involves a rhetorical question format rooted in Taiwanese Internet slang culture, commonly known as ``\CJKtext{問卦}'' (gossip-style inquiry), which appears in roughly 69\% of the false negatives. Although framed as casual questions, these questions rarely function as genuine information-seeking requests. Instead, they serve as an indirect vehicle for harmful intent — soliciting offensive opinions, embedding unsafe instructions in interrogative form, softening identity-targeted harassment, or recasting harmful content to evade explicit detection. This creates a surface-form/pragmatic-intent mismatch: the content appears benign under literal reading, yet its communicative purpose remains unsafe. Because such discourse conventions are characteristic of Taiwanese Internet slang but underrepresented in common Chinese corpora, guardrail models may lack the cultural-linguistic grounding needed to reliably detect them.

Manual inspection further shows that many missed samples contain implicit hate, sexual content, or culturally specific slang commonly used in Taiwan. For example, as shown in Case 1 of Table~\ref{tab:qualitative_examples}, Llama-Guard-3-8B fails to recognize identity-based hostility when the harmful intent is expressed implicitly rather than using explicit violent language. Although the prompt does not directly advocate violence, its phrasing implies the disappearance of a protected identity group and thus reflects identity-based hate. 

We also observe that sexually unsafe content is often missed when it is conveyed through colloquial or indirect expressions rather than standard explicit vocabulary. As illustrated in Case 2 of Table~\ref{tab:qualitative_examples}, the term ``\CJKtext{修*}'' shows an explicit sexual connotation but the prompt was misclassified as benign. This failure is exacerbated by linguistic code-mixing, as the term represents a Taiwanese Hokkien expression transcribed into Han characters via phonetic borrowing, rather than a standard Mandarin lexicon. Semantically, the expression is equivalent to the Mandarin term for sexual intercourse; however, its non-standard orthography and localized usage impede the ability of general-purpose safety models to achieve consistent recognition.

These findings suggest that Llama-Guard-3-8B's errors are not random but are systematically related to localized linguistic characteristics, including indirect rhetorical framing, implicit hate, and Taiwanese colloquial slang. This highlights the importance of incorporating region-specific linguistic patterns when constructing safety datasets for guardrails.

\subsubsection{What Qwen3Guard Missed}

We analyze the error distribution of Qwen3Guard using the same strategy. Compared to other baselines, Qwen3Guard (strict) achieves relatively strong overall performance with high recall. However, its FPR is noticeably higher than TWGuard's (0.039 vs. 0.002). We therefore focus our analysis on these false positive cases.

Among the 21 false positive samples, manual inspection reveals that 18 involve politically related topics, accounting for the majority of the observed false positives. A closer examination shows that many of these prompts involve general discussions of public figures or politically adjacent topics and do not have harmful or manipulative intent in the Taiwanese linguistic context. Case 3 in Table~\ref{tab:qualitative_examples} provides a representative example: the prompt asks the model to impersonate the President of Taiwan and provide a neutral analysis of the Russia--Ukraine war and its implications for cross-strait relations. Although it is politically sensitive in some Mandarin-speaking regions, it falls under legitimate civic discourse in our annotation process and is therefore labeled safe. Qwen3Guard nevertheless predicts unsafe, suggesting a broader political sensitivity threshold.

While such sensitivity may help maintain high recall, it can also lead to more false positives in environments where political topics are discussed in everyday conversations. In the Taiwanese linguistic context, conversations about public figures, social issues, and cross-strait topics often take place in casual settings. As a result, overly broad political filtering may lead to unnecessary blocking in localized discourse settings.

\subsubsection{Findings}

Overall, the two baseline models exhibit distinct qualitative error patterns under our evaluation. Llama-Guard-3-8B primarily suffers from false negatives in cases involving Taiwanese slang, indirect expressions, implicit hate, and online discourse patterns common in local online communities. In contrast, Qwen3Guard demonstrates strong recall but produces more false positives due to broader sensitivity to politically related topics. These observations suggest that safety performance in guard models can be strongly influenced by the coverage of region-specific linguistic and cultural patterns in training data. Incorporating Taiwan-contextualized data can therefore help improve safety boundary calibration in real-world Taiwanese environments.

\subsection{Ablation Study}

We investigate the robustness of TWGuard by systematically varying two key training factors: class distribution and data augmentation, to understand their individual contributions to model behavior. Table~\ref{tab:ablation_split_settings} summarizes the training split settings used in our ablation experiments.

\begin{table}[t]
\caption{Training split settings for the ablation study.}
\label{tab:ablation_split_settings}
\centering
\begin{tabular}{lcc}
\hline
\textbf{Setting} & \textbf{\# of samples} & \textbf{Positive rate} \\
\hline
TWGuard & 16,267 & 7.18\% \\
TWGuard (balanced with aug.) & 2,336 & 50\% \\
TWGuard (balanced w/o aug.) & 1,078 & 49.6\% \\
\hline
\end{tabular}
\end{table}

\begin{figure}[t]
\centering
\includegraphics[width=0.8\linewidth]{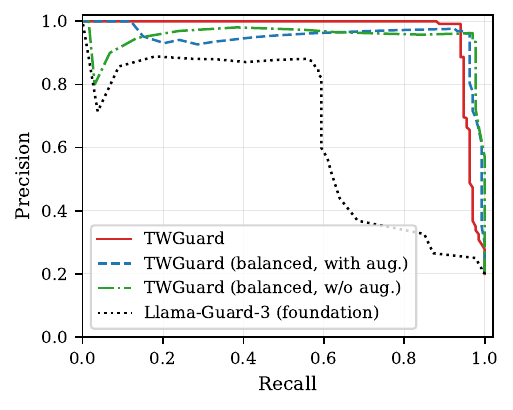}
\caption{
Precision--recall curves under different training split settings.
}
\label{prcurve_ft_3versions}
\end{figure}

\subsubsection{Performance under Different Training Data Distributions}
Since TWGuard is trained on data from the finalized data pool, which preserves the natural class imbalance of real-world Taiwan-specific data (a 0.072 unsafe rate), we further conduct an additional controlled comparison to isolate the effect of class distribution.

As shown in Table~\ref{tab:ablation_distribution}, TWGuard trained under a naturally imbalanced distribution achieves higher precision, lower FPR, and a slightly better F1-score than the balanced variant, although recall is slightly lower. These results suggest that preserving the original class distribution helps the model better control false positives while retaining competitive unsafe detection capability.

\begin{table}[t]
\caption{Effect of training data distribution.}
\label{tab:ablation_distribution}
\centering
\begin{tabular}{lccccc}
\hline
\textbf{Model} & \textbf{Precision} & \textbf{Recall} & \textbf{F1-score} & \textbf{FPR} & \textbf{AUPRC} \\
\hline
TWGuard (balanced) & 0.962 & 0.962 & 0.962 & 0.009 & 0.950 \\
TWGuard & 0.992 & 0.940 & 0.965 & 0.002 & 0.970 \\
$\Delta$ (TWGuard $-$ balanced) & +0.030 & -0.023 & +0.003 & -0.007 & +0.019 \\
\hline
\end{tabular}
\end{table}

Figure~\ref{prcurve_ft_3versions} further shows that this difference is visible across a broad operating range. Compared with the balanced setting, the model trained under the naturally imbalanced distribution maintains higher precision across most recall levels. This indicates that, as the threshold is relaxed to recover more unsafe cases, the imbalanced model introduces fewer additional false positives. In our setting, this behavior is consistent with the natural class distribution more closely matching the composition of real Taiwan-specific user data, in which unsafe content constitutes only a small fraction of the data. Training under this distribution may therefore help the model avoid over-predicting unsafe labels, leading to more reliable unsafe content detection.

Consistent with this observation, the AUPRC of the imbalanced model is also slightly higher than that of the balanced model, indicating improved overall ranking quality under the natural data distribution.

\subsubsection{Effect of Data Augmentation}

To further understand the contribution of data augmentation, we build on the balanced training setting introduced in the previous ablation and compare two balanced variants that differ in whether augmented samples are included. This controlled comparison allows us to isolate the effect of augmentation without the influence of class imbalance.

As shown in Table~\ref{tab:ablation_augmentation}, the model trained without augmented data achieves a precision of 0.935, recall of 0.977, F1-score of 0.956, and an FPR of 0.017. After augmented samples are incorporated, precision increases to 0.962, recall becomes 0.962, F1-score improves to 0.962, and FPR decreases to 0.009.

\begin{table}[t]
\caption{Effect of data augmentation under balanced training.}
\label{tab:ablation_augmentation}
\centering
\begin{tabular}{lccccc}
\hline
\textbf{Model} & \textbf{Precision} & \textbf{Recall} & \textbf{F1-score} & \textbf{FPR} & \textbf{AUPRC} \\
\hline
TWGuard (balanced, w/o aug.) & 0.935 & 0.977 & 0.956 & 0.017 & 0.952 \\
TWGuard (balanced, with aug.) & 0.962 & 0.962 & 0.962 & 0.009 & 0.950 \\
$\Delta$ (with $-$ w/o) & +0.027 & -0.015 & +0.006 & -0.008 & -0.002 \\
\hline
\end{tabular}
\end{table}

The results indicate that augmentation reduces false positives while maintaining a strong unsafe detection capability. The gain in precision, together with only a modest decrease in recall, results in an overall improvement in F1-score. The lower false positive rate further suggests that augmented samples help the model better distinguish between safe and unsafe content, reducing over-triggering without substantially increasing the number of missed unsafe cases.

In the balanced-training comparison shown in Figure~\ref{prcurve_ft_3versions}, the two models exhibit broadly similar PR curves across the entire recall range, indicating that data augmentation does not substantially change the model’s overall ranking ability. This observation is consistent with the nearly identical AUPRC values of the two models.

In the upper-left region of the PR curves, both models show a small dip in precision, suggesting that a few safe samples are ranked highly among the top-scored unsafe predictions. After augmentation, this dip becomes slightly smoother and less abrupt, indicating that augmented data partially stabilizes the ranking of the highest-scoring predictions.

These observations indicate that augmentation mainly refines the model's behavior locally along the PR curve, rather than substantially changing its global ranking quality. This localized improvement aligns with our data augmentation design, which aims to enrich training coverage near the safe–unsafe boundary and expose the model to more difficult input variants.

\subsubsection{Findings}

Overall, the ablation results show that the performance differences across training variants are relatively small, indicating that the proposed approach is robust across multiple training settings rather than dependent on a single specific recipe. Across the ablation settings, the models consistently maintain strong precision, recall, and F1-score, suggesting that the primary benefit comes from the localized data pipeline, while augmentation and class-distribution choices mainly affect the precision–recall trade-off and false-positive behavior.

False positive control is the most practically meaningful difference among these variants. The naturally imbalanced setting on which TWGuard is trained achieves the lowest FPR of 0.002, compared with 0.009 for the balanced setting with augmentation and 0.017 for the balanced setting without augmentation. This corresponds to an approximately 5x reduction relative to the augmented balanced variant and nearly 9x relative to the non-augmented balanced variant. Overall, these comparisons suggest that training configurations better aligned with real-world data characteristics can help the model avoid over-predicting unsafe labels. This is especially important in moderation scenarios where safe inputs greatly outnumber unsafe ones. The results further suggest that the data dependency can be further reduced (from 16K to 1K) while maintaining a manageable FPR in our approach.

Overall, our localized optimization approach remains effective even when the training setup is adjusted according to practical trade-offs between detection performance, false-positive control, and training cost, supporting the enhanced applicability of the proposed methodology beyond TWGuard as a single case study.

\subsection{Capability Retention}

To further evaluate whether Taiwan-specific fine-tuning weakens the base model’s capability, we additionally test our model on ToxicChat~\cite{toxicchat}, an in-the-wild safety benchmark. 

On this benchmark, our fine-tuned model achieves an F1 score of 0.645, higher than the 0.538 reported for the foundation model in ~\cite{qwen3guard}. The result suggests that Taiwan-specific fine-tuning does not degrade the model’s general moderation capability on an out-of-distribution dataset. Our hypothesis is that both our Taiwan-specific corpus and ToxicChat are derived from real-world data. Although the two datasets differ in language, they may share similar characteristics found in real-world unsafe cases, such as informal wording, ambiguous intent, and jailbreak-like expressions. This similarity may partially explain why our model remains competitive and even performs slightly better in the experiment.

\subsection{Why Not Just Red-teaming?}
The most straightforward approach to demonstrate the risks and effectiveness of cultural diversity in safety alignment is to manually curate adversarial attack samples covering sensitive topics such as national identity and religion, themes frequently observed in Sovereign AI and red-teaming research~\cite{yin2024safeworld,hu2025toxicity,aitaiwansovbenchmark}.

However, we deliberately opted for a data-driven approach using organic, real-world collections rather than synthetic or manually engineered templates. This methodological choice aims to prevent the over-amplification of specific high-profile scenarios, ensuring that our evaluation metrics authentically reflect actual user experiences and linguistic nuances within the local context. By giving precedence to ecological validity over isolated stress-testing, we provide a more representative assessment of how guardrails perform against the diverse and unpredictable nature of real-world interactions. 

The results shown in Figure~\ref{fig:results} provide some evidence that our dataset does not focus solely on such corner cases, as baselines such as ShieldGemma also perform at an acceptable level, even though it is not trained to recognize risks in Taiwanese culture or in Mandarin.

\subsection{Limitations}
Several limitations are worth noting in this study. First, the proposed approach is validated as a proof-of-concept in the Taiwan cultural context; while the culture-aware fine-tuning strategy is generalizable in principle, its effectiveness in other linguistic and cultural settings remains unvalidated. To encourage broader exploration, we release our model weights so that researchers can extend this approach to cultural contexts of their interest. Second, the fine-tuning dataset, while curated to be culturally representative, may not capture the full diversity of harmful content expressions within Taiwanese society; broader community participation in data collection could improve coverage. Third, this work focuses exclusively on text-based prompts and does not address multimodal inputs or model responses, which represent important directions for future extension. Fourth, due to data scarcity in this domain, cross-dataset evaluation was not feasible; we attempted to mitigate this by re-splitting the current dataset and plan to collect additional data to enable more rigorous evaluation. Finally, our baseline comparisons focus on open-weight models to ensure reproducibility; comparisons with proprietary systems such as the OpenAI Moderation API or Google Perspective API remain an open item. We acknowledge these aspects as opportunities for future work.

\section{Conclusion}

This study addresses a critical gap in current AI safety frameworks, namely the lack of linguistic sensitivity and rigorous localized evaluation in existing guardrail models. We propose a culture-aware optimization paradigm that focuses on high-quality, culturally representative datasets over the conventional approach of large-scale multilingual expansion.

The implementation of our proposed approach, instantiated as TWGuard, demonstrates significant performance gains within the Taiwan linguistic context. Experimental results show that TWGuard achieves a huge improvement in F1 score (+0.289) compared to its foundation model. Notably, the model attained an FPR of 0.002, representing a 94.9\% reduction over the Qwen3Guard in strict mode (0.039). Furthermore, our work demonstrates that safety alignment is highly feasible even with limited linguistic data and computational power, both of which are essential factors for practical deployment. This feasibility is further supported by our Safe-Case-Guided Augmentation strategy, which improves coverage of unsafe cases under limited data conditions. By leveraging 16,267 training samples, which is less than half of the Mandarin dataset size used in NemoGuard and Qwen3Guard, our approach achieves superior results. The results from our ablation studies suggest that data dependency can be further reduced while maintaining a manageable FPR. Model weights will be released to ensure reproducibility.

Together, this work lays a foundation for regional communities to establish AI safety standards by optimizing guardrail models for their own linguistic contexts, rather than accepting boundaries imposed by dominant languages. The inadequacy of the latter is reconfirmed by our findings, demonstrating their potential sensitivity to linguistic environments. We advocate for a transition toward culture-based categorization as a more principled and effective framework for ensuring AI safety and compliance across different geopolitical and social landscapes.

\section*{Acknowledgment}
The authors thank B.-S. Ke and Y.-T. Chung for their contributions to dataset annotation and P.-C. Chen for his support in the project and in coordinating resources. The authors also thank researchers in the Cloud Computing Laboratory and Telecommunication Laboratories for their support in all aspects. Finally, the authors thank Meta for their open-weight model and open source toolchain as our research foundation.

\end{document}